\magnification=\magstep1
\baselineskip=20pt
\centerline{\bf Temperature Dependence of the Index of Refraction of Fused
Silica}
\centerline{\bf Answer to Question \# 50}
\bigskip
\centerline{J. I. Katz}
\centerline{Department of Physics and McDonnell Center for the Space Sciences}
\centerline{Washington University, St. Louis, Mo. 63130}
\bigskip
Because the coefficient of thermal expansion of fused silica$^1$ is so small 
($\approx 0.4 \times 10^{-6\,\circ}$K$^{-1}$ at room temperature), it is 
evident that the variation in refractive index$^2$ ($dn/dT \approx 1.1 \times 
10^{-5}\,^\circ$K$^{-1}$) must result from a change in the 
internal structure of the material with temperature.  Note, also, that 
dilution of dispersive material by thermal expansion would tend to produce a
negative $dn/dT$, opposite to what is observed.  Here we develop the most 
elementary possible model.
It is qualitatively, but not quantitatively, consistent with the data.
    
We write the complex dielectric constant $\epsilon(\omega)$ as the sum of real
and imaginary parts
$$\epsilon(\omega) = \epsilon^\prime(\omega) + i \epsilon^{\prime\prime}
( \omega ), \eqno(1)$$
with $\epsilon^\prime(\omega)$ and $\epsilon^{\prime\prime}(\omega)$ each real.
Then we use one of the Kramers-Kronig relations$^3$
$$\epsilon^\prime(\omega) = 1 + {2 \over \pi} {\rm P} \int^\infty_0 {s
\epsilon^{\prime\prime}(s) \over s^2 - \omega^2}\,ds. \eqno(2)$$
We assume that $\epsilon^{\prime\prime}(\omega)$ is narrowly peaked near a high
(ultraviolet) freauency $s_0 \gg \omega$.
    
At zero temperature $\epsilon^{\prime\prime}(\omega)$ may be a Dirac
$\delta$-function, or a sum of $\delta$-functions, each corresponding to an
allowed vibronic transition, or to its analogue in an amorphous solid. At
finite temperature $T$ the excited vibrational (phonon) substates of the ground
electronic state are excited. In general, these have the same allowed
transitions to excited electronic states as the vibrational ground state, and
we will assume they have the same oscillator strengths. Assuming a classical
distribution of ground state vibrational excitation (requiring, in a crystal,
a temperature in excess of the Debye temperature) yields a simple analytical
model for $\epsilon^{\prime\prime}(\omega)$:
$$\epsilon^{\prime\prime}(s) = \cases{{\displaystyle C \hbar \over \displaystyle
k_B T}\exp[\hbar (s - s_0)/k_B T] & $s < s_0$ \cr 0 & $ s \ge s_0$,\cr}
\eqno(3)$$
where $C$ is a normalization constant. This model assumes a single allowed
electronic transition of (zero-temperature) frequency $s_0$, but may be applied
if there are many such transitions, each with a frequency $\gg \omega$, if $s_0$
is taken as an oscillator strength-weighted average. It tacitly neglects any
temperature dependence of the Franck-Condon factors.
    
Using (3) in (2) yields
$$\epsilon^\prime(\omega) = 1 + {2 \over \pi} {\rm P} \int^{s_0}_0 {s \over
s^2 - \omega^2} {C \hbar \over k_B T} \exp [\hbar (s - s_0) / k_B T]\,ds
\eqno(4)$$
If we approximate $k_B T \ll \hbar s_0$ then the integrand is narrowly peaked
around $s = s_0$. The lower limit of the integral may be extended to $-\infty$
and $s/(s^2 - \omega^2)$ expanded in a Taylor series about $s_0$:
$${s \over s^2 - \omega^2} = {1 \over s_0} - {(s - s_0) \over s_0^2} + {1 \over
2} {4 \omega^2 \over s_0^5} (s - s_0)^2 - \ldots. \eqno(5)$$
This removes the pole in the integrand. Then the integral is elementary,
yielding
$$\epsilon^\prime(\omega) - 1 = {2 \over \pi} {C \over s_0}, \eqno(6)$$
which determines $C$.
   
Differentiating (4) yields
$${d \epsilon^\prime(\omega) \over dT} = - {2 \over \pi} \int^{s_0}_0 {s \over
s^2 - \omega^2} {C \hbar \over k_B T} \left({\hbar (s - s_0) \over k_B T} + 1
\right) \exp [\hbar (s - s_0)/k_B T]\,ds. \eqno(7)$$
Again taking $k_B T \ll \hbar s_0$ and $\omega \ll s_0$, the first term in the
expansion (5) integrates to zero but the second yields the result
$${d \epsilon^\prime(\omega) \over dT} = {2 \over \pi} {C k_B \over \hbar s_0^2}
= [\epsilon^\prime(\omega) - 1]{k_B \over \hbar s_0}. \eqno(8)$$

In terms of the refractive index $n = (\epsilon \mu)^{1/2} = \epsilon^{1/2}$
$${d n(\omega) \over dT} = {n^2 - 1 \over 2n} {k_B \over \hbar s_0}. \eqno(9)$$
In order to evaluate this numerically we need to estimate $s_0$. Fused silica
has very strong ultraviolet absorption centered around a wavelength of 1200 \AA,
implying $\hbar s_0 \approx 1.6 \times 10^{-11}$ erg. Then, with $n = 1.445$
in the visible and near-infrared, we have
$${dn \over dT} = 3.2 \times 10^{-6\,\circ}{\rm K}^{-1}. \eqno(10)$$
   
This theoretical value is about 3.4 times less than the measured$^2$ value of
$1.1 \times 10^{-5\,\circ}$K$^{-1}$. The most plausible explanation of the 
discrepancy is that the Franck-Condon factors are increasing functions of 
temperature, as increasing vibrational excitation of the electronic ground 
state increases the range of internuclear separations, providing a better 
match to upper states in which the equilibrium internuclear separation is 
greater. This effect cannot be calculated analytically.
    
The theory predicts that $dn/dT$ be nearly independent of frequency because 
$n$ is nearly independent of frequency.  This is observed$^2$ to be true to 
within about 10\% for wavelengths between 3000 \AA\ and 4$\,\mu$. $dn/dT$ is
measured to increase with decreasing wavelength for $\lambda < 3000$ \AA, 
where the assumption $\omega \ll s_0$ is no longer valid.  The theory could 
readily be extended to these larger values of $\omega$ by means of a 
straightforward numerical integration.
\bigskip
\parindent=0pt
$^1$ American Institute of Physics {\it American Institute of Physics
Handbook} (McGraw-Hill, New York, 1957). \par
$^2$ I. H. Malitson, ``Interspecimen Comparison of the Refractive Index of
Fused Silica,'' J. Opt. Soc. Am. {\bf 55}, 1205--1209 (1965). \par
$^3$ J. D. Jackson {\it Classical Electrodynamics} (Wiley, New York, 1962),
Prob. 7.9. \par
\vfil
\end